\documentclass[conference]{IEEEtran}

\usepackage{amssymb}
\usepackage{graphicx}
\usepackage{amsmath}
\usepackage{array}
\usepackage{float}
\begin{document}
\title{A Probabilistic Approach for Data Management in Pervasive Computing Applications}

\author{\IEEEauthorblockN{Kostas Kolomvatsos}
\IEEEauthorblockA{Department of Informatics and Telecommunications\\ University of Thessaly\\
Papasiopoulou 2-4, 35131, Lamia, Greece\\
e-mail: kostasks@uth.gr
}\\
}

\maketitle

\begin{abstract}
Current advances in Pervasive Computing (PC) 
involve the adoption of the huge infrastructures of the Internet of Things (IoT) and the Edge Computing (EC).
Both, IoT and EC, can support innovative applications around end users
to facilitate their activities. 
Such applications are built upon the collected data and the appropriate 
processing demanded in the form of requests. 
To limit the latency, instead of relying on Cloud for data storage and processing, 
the research community provides a number of models for data 
management at the EC. 
Requests, usually defined in the form of tasks or queries, demand 
the processing of specific data. 
A model for pre-processing the data preparing them and detecting 
their statistics before requests arrive is necessary.
In this paper, we propose a promising 
and easy to implement scheme for selecting 
the appropriate host of the incoming data 
based on a probabilistic approach. Our aim is to 
store similar data in the same distributed datasets to have, beforehand, 
knowledge on their statistics while keeping their solidity at high levels.
As solidity, we consider the limited statistical deviation of data, thus, 
we can support the storage of highly correlated data 
in the same dataset. 
Additionally, we propose an aggregation mechanism
for outliers detection applied just after the 
arrival of data. 
Outliers are transferred to Cloud for further processing. 
When data are accepted to be locally stored, we propose a model for selecting the appropriate 
datasets where they will be replicated for building a fault tolerant system.
We analytically describe our model and evaluate it through extensive 
simulations presenting its pros and cons.   
\end{abstract}

\begin{IEEEkeywords}
Pervasive Computing, Internet of Things, Edge Computing, Data Storage, Accuracy, Probabilistic Model
\end{IEEEkeywords}

\IEEEpeerreviewmaketitle

\section{Introduction}
The combination of the Internet of Things (IoT) and Edge Computing (EC)
provides a promising infrastructure for supporting innovative 
Pervasive Computing (PC) applications.
The aim of PC is to offer ambient intelligence to end users having 
devices and services interacting with them. 
IoT devices are, usually, carried by end users or they are present in 
the environment in close distance with them facilitating the envisioned interactions and the collection of data.
Then, data become the subject of processing activities to create knowledge and support novel applications.
At the EC, we can detect numerous nodes that are,
in an upwards mode, connected with Cloud
for transferring data and ask for more advanced processing. 
A new trend is the collection and storage of data
at the EC to eliminate the latency in the provision 
of responses in various requests (instead of always relying on Cloud).
EC nodes have direct connection with IoT devices and become the hosts
of a high number of geo-distributed datasets. 
Obviously, as new data arrive, one can observe the `evolution' of the local datasets 
as depicted by their statistics.
This evolution is represented by updates in the statistical information, 
e.g., the mean and standard deviation may be altered as new information is 
retrieved by the IoT devices.

Users/applications perform requests upon the distributed datasets to create 
knowledge or ask for analytics.
Requests can have the form of tasks 
(e.g., apply a machine learning model and report the outcome) or
queries (e.g., report the list of data that meet a specific condition).
When a request is set, the `base case' model is to 
launch it across the network and search the information
that end users/applications are interested in \cite{xa}. 
However, this involves an increased messaging overhead paid without any reason
for nodes/datasets that do not match to the desired conditions set by the request.
The most efficient solution is to have a view, beforehand, on the statistics of 
the available data and decide upon the matching between the requests and the
distributed datasets. Through this approach, we can eliminate the cost of allocating tasks/queries to datasets that do not match to the defined conditions as their execution will return an empty set. A number of research efforts propose techniques for the 
optimal tasks/queries allocation into a number of processing nodes, e.g.,
\cite{dong}, \cite{gu}, \cite{josilo}, \cite{karanika}, \cite{kolomvatsosapplied}, \cite{kolomvatsoscomputing}, \cite{kolomvatsosfgcs}.

Another challenge is to keep 
the consistency and accuracy of data 
at high levels as the critical 
statistical information that depicts the 
quality of the collected data \cite{refcong}.
Accuracy refers to the closeness of estimates 
to the (unknown) exact or true values \cite{refec}, i.e.,
it depicts the error between the 
observation and the real data.
Envision a new data vector arriving from an IoT device to an 
EC node. The discussed vector may affect
the accuracy of the local dataset as it 
may not match to dataset's current statistics
(e.g., it may be an outlier to the specific dataset).
We borrow the concept of 
`solidity' to represent 
the closeness of data into a dataset \cite{kolomvatsosspringer} and adopt it
to build a model for the efficient allocation of data to the 
appropriate datasets.
A solid dataset exhibits a high accuracy 
realized when the error/difference between the involved data is low, 
e.g., the
standard deviation may be be limited. 
We have to notice that accuracy 
is significant as efficient response plans,
for each type of tasks/queries, may be defined.
Apart from maintaining the solidity of datasets, we have to focus 
on data replication to support a fault tolerant EC infrastructure. 
Such an approach will provide benefits when connectivity is limited and
data can be processed to deliver the required responses.
In the discussed ecosystem, there is no need for additional data migration
that will burden the network.
It is preferable to migrate tasks/queries instead of circulating huge volumes of data.
Replication in combination with the distributed data storage 
may assist in the elimination of the probability of data loss and cope
with IoT nodes failure as well \cite{amrutha}.

This paper targets to support the distributed nature of the EC ecosystem
and the provision of an ensemble model
for outliers detection 
combined with methodologies for the selection of the appropriate 
datasets where new data vectors should be stored.
EC nodes adopt a monitoring scheme for examining the incoming data
and a decision making model for performing the envisioned allocations.
For outliers detection, we rely on simple, however, fast ensemble approach 
extending our previous efforts in the domain
\cite{kolomvatsosscalcom}.
Instead of adopting an individual majority voting method upon multiple 
outlier indicators, we study a double majority scheme to conclude if the incoming 
data are outliers. 
Furthermore, when data are not considered as outliers, we 
study the time series of difference `quanta',
i.e., the time series of the difference between the incoming data
and the synopses of the available datasets as reported by EC nodes.
We consider that nodes at pre-defined epochs share the synopses of their
datasets to become the basis for the 
proposed replication process. 
Evidently, these quanta show the statistical `behaviour' of data synopses
exhibiting the trends in every dataset.
The new data are placed at the datasets where 
the similarity with the synopses is high, i.e., the difference quanta are limited.
In any case, our replication model is based on historical quanta 
realizations instead of relying to the latest one.
Compared with our previous efforts \cite{kolomvatsosspringer},
\cite{kolomvatsosscalcom}, the proposed model exhibits the following differences:
(i) we provide an aggregation mechanism for the management of multiple outlier indicators upon multiple datasets instead of using a limited number of indicator functions
like in \cite{kolomvatsosspringer}, \cite{kolomvatsosscalcom}. The proposed scheme
can be applied upon any number and any type of indicators;
(ii) for the replication process, we rely on a probabilistic approach upon multiple historical quanta instead of using an uncertainty driven model \cite{kolomvatsosspringer}.
The adoption of the uncertainty management scheme requires 
the manual definition of a rule base which is not necessary in the current model. 
The following list reports on the contributions of our paper:
\begin{itemize}
	\item we provide an ensemble model for the aggregation of multiple outlier indicators
	upon a double majority voting scheme;
	\item we support the data replication process with a probabilistic model based on multiple historical synopses reported by EC nodes. Our aim is to detect the similarity of the incoming data with the available datasets upon their past trends;
	\item we present the outcomes of an extensive set of experiments to reveal the characteristics of the proposed approach.
\end{itemize}

The paper is organized as follows. Section \ref{related} reports on the 
related work in the domain.
In Section \ref{overview}, we present the preliminary information while in Section
\ref{mechanism}, we provide the analytical description of our model.
In Section \ref{setup}, we present our experimental evaluation 
and in Section \ref{conclusions},
we conclude our paper by providing our future research directions.

\section{Related Work}
\label{related}
Outliers detection and management is a widely studied research domain.
The interested reader can refer in \cite{zhang} for an extensive survey for getting insights on the proposed
methodologies. 
These methods target to identify objects deviating from a group of other objects.
Deviating objects depict an 
abnormal behaviour compared to the natural evolution of the collected 
data. 
If we compare univariate with multivariate data, one can argue that the latter 
scenario is more prone to outliers, i.e., 
the effects of outliers in the statistics of data have higher impact than in the 
former case \cite{meade}. 
Recall that 
in the vast majority of the application domains,
data are reported in a multivariate `form',
i.e., tuples/vectors
making the detection of outliers a difficult task \cite{leys}.  
For detecting outliers, we have to rely on statistical metrics
like the covariance matrix \cite{filzmoser}. 
Mahalanobis distance is the most representative metric that builds upon the 
covariance matrix adopted to detect the correlations between the multiple dimensions of
data vectors \cite{mahalanobis}.
Other techniques involve the Cook's distance \cite{cook} and the leverage model \cite{leverage}. 
The former metric
estimates variations in regression coefficients after removing
each observation, one by one.  
The latter model performs in a similar way to the 
Mahalanobis distance as it is
based on the sstudy of residuals and their distance from the mean vector.
Additional techniques can be found in the relevant literature like the $\chi^{2}$ metric 
for identifying deviations from the 
multidimensional normality.
An extension of the Mahalanobis distance 
is proposed in \cite{leys}, i.e., a variant based on the minimum
covariance determinant, a more robust process that is easy to implement. 
In \cite{baxter}, the interested reader can find a comparison of outlier detection methods.

Data replication is usually adopted to achieve two goals, 
i.e., the minimization of the latency in the provision of responses and the support
of fault tolerant systems.
In the first axis, we avoid data or requests migration while in the second
axis, we are able to perform the desired processing
even if a node is not available.
The replication of data is a technique 
adopted in Wireless Sensor Networks (WSNs) where we need to 
upload IoT
data from a set of sensor gateways 
on distributed Cloud storage \cite{kumar}.
In this scenario, we can consider multiple mini-Clouds 
as the hosts of 
data 
taking replication requirements for each data item into
account.
A distributed algorithm for the replication placement is proposed in \cite{zaman}.
The authors propose the use of distributed storages and a method 
for improving
the efficiency of objects replication. 
The main subject of \cite{venkatesan} is the reliability assessment
of clustered and declustered
replica placements. 
In
\cite{bonvin}, the authors present a self managed key-value store which dynamically
allocates the resources of a data Cloud to several
applications, thus, it maintains differentiated availability guarantee
for different application requirements.

Other efforts formulate the problem around 
what, when and where data replication should take place \cite{dadria}.
Such a modelling can assist in applying optimization schemes 
for concluding the best possible action.
Data replication may also assist in eliminating the need for 
migrating huge volumes of data as bulk data
transfer protocols aim to do \cite{ren}. 
We can perform a selective
replication under constraints avoiding to circulate all the collected data in the network.
Such a selective approach can be realized online
\cite{tai},
however, 
under the goal of choosing the 
appropriate hosts to have data close to the appropriate users.
In networks, where nodes exhibit limited energy resources, 
we have to balance the number of replicas to the energy consumption \cite{boru}. 
We can achieve the discussed goal 
adopting compression and reduction techniques before the 
replication takes place. 
This step may be adopted in the pre-processing stage where
data are pre-processed before sending them to the
storage node to reduce the energy consumption without
affecting the data quality requirements \cite{cappiello}.
The disadvantage of the discussed approaches is that they require 
a complex process and increased computational resources 
for compression/decompression. 
Modelling the energy footprint of nodes may be also adopted to 
create replication plans that aim to reduce the 
energy consumption \cite{takahashi}.
In \cite{luo}, the authors incorporate a load
balancing approach when performing the desired replication process.
They assume that every node has a local memory adopted
to store neighbour nodes memory contents. 
ProFlex
\cite{maia} is proposed to deal with the communication requirements
for transferring data from `weak' to 
strong nodes. 
TinyDSM \cite{piotrowski} exhibits a reactive replication method
that distributes replicas based on a random strategy 
according to the number and
the density of replications.
Finally, in \cite{rajabi}, 
the authors propose 
a low-complexity distributed 
mechanism to increase the capacity of WSN-based
distributed storage, optimizing communication and
decreasing energy usage. 
Data are collected periodically by the sink node
being removed from nodes' memories while, based on a greedy distribution
storage scheme, each node reports its memory
condition to other nodes.

\section{Problem Description \& Preliminaries}
\label{overview}
The problem under consideration involves a set of
IoT devices reporting data to an EC node
being member of a group of $N$ nodes, i.e., 
$n_{1}, n_{2}, \ldots, n_{N}$.
EC nodes receive the data 
and perform an initial pre-processing for deciding 
their storage or rejection.
A rejection corresponds to the transfer of 
data to Cloud while a decision related to storage fires
another round of processing to deliver the appropriate 
node where the incoming data should be replicated.
EC nodes formulate local datasets,
i.e., $\left\lbrace D_{1}, D_{2}, \ldots, D_{N}\right\rbrace$ upon which 
they perform the requested processing activities.
The discussed datasets are updated upon the arrival
of new multivariate data vectors, i.e., $\mathbf{x}=[x_{1}, x_{2}, \ldots, x_{M}]$.
Additionally, due to the proposed replication process, data vectors
can be exchanged by peer nodes as the result of the proposed decision mechanism. 
$D_{i}$s exhibit specific statistical 
information and data synopses can be extracted upon them 
to be disseminated 
in peers for supporting the envisioned decision making.
For instance, the synopsis for the $i$th dataset 
can refer in a vectorial space via a $M$-dimensional vector,
i.e., $s_{i}=[s_{i1}, s_{i2}, \ldots, s_{iM}]$ conveying statistical information for each of the adopted 
$M$ dimensions. 
Synopses
may represent linear multivariate regression coefficients, clusters 
centroids (in a clustering scheme), the first Principal Components (PCs) (in linear data compression)
or very simple statistical measures like the mean or standard deviation. 
We target to keep EC nodes informed about the data present to their peers, thus,
support decision making upon fresh information for the collected data.
Obviously, there is a trade off between the frequent data synopses calculation in the burden of 
network performance (a high number of messages is required to transfer synopses to peers) compared
to a less frequent synopses extraction in the burden of decision making upon `obsolete'
data synopses.
The study of the discussed trade off is beyond the scope of this paper.

Initially, every node $n_{i}$ should 
detect if $\mathbf{x}$ is an outlier not only 
compared with the local dataset $\mathbf{D}_{i}$ but also 
with the remaining repositories present at peer nodes.
Our aim is to detect if $\mathbf{x}$ significantly 
deviates from the `ecosystem' of datasets, thus, 
no dataset could become the host of $\mathbf{x}$. 
For this, we rely on an ensemble approach and 
study the involvement of multiple outlier detection methods. 
$\mathbf{x}$ is rejected when multiple indicators depict an outlier judgement 
in multiple datasets.
The discussed synopses `participate' in the delivery 
of the outlier indication as represented by the 
indicators outcome. 
We consider that $V$ outliers indicator functions
are available, i.e., 
\begin{equation*}
I_{j}(\mathbf{x}) = \begin{cases}
             1  & \text{if } \mathbf{x} \text{ is an outlier with probability } \beta_{j} \\
             0  & \text{Otherwise } 
       \end{cases}
\end{equation*} 
for all $j \in \left\lbrace 1,2, \ldots, V \right\rbrace$.

Having calculated $V$ indicators for a dataset 
$D_{i}$, we propose a  
consensus model for deciding if $\mathbf{x}$
is an outlier or not.
Our aim is to check if $\mathbf{x}$ is an outlier for 
a high number of datasets, thus, it significantly deviates
from the data ecosystem currently present at the EC. 
The final decision is based on a double majority voting model,
i.e.,
an ensemble scheme taking into consideration 
not only an individual dataset by the entire ecosystem. 
When $\mathbf{x}$ is not an outlier, 
$\mathbf{x}$ is stored to the node $n_{i}$, where it is initially reported,
if it is highly correlated with $D_{i}$.
Apart from the local dataset, it should be also replicated to 
peers exhibiting a significant correlation with $\mathbf{x}$.
The correlation is detected upon the latest $W$
synopses reports and the similarity between them and 
$\mathbf{x}$.
We define the concept of the 
\textit{similarity quantum}, i.e., 
the magnitude of the similarity (as exposed by the 
numeric difference) between $\mathbf{x}$ 
and every synopsis reported by peers.
Upon the latest $W$ similarity/difference quanta,
we expose their distribution and deliver the probability 
of having quanta upon a threshold.
This probability is adopted to rank the available datasets and 
decide where $\mathbf{x}$ will be replicated.


\section{The Proposed Mechanism}
\label{mechanism}

\subsection{Probabilistic Outliers Detection}
\label{outliers}
The preliminaries section defines the basis for our 
outliers detection model upon $V$ outliers detection schemes 
and $N$ datasets.
We focus on the `collection' of our indicators and define a two-dimensional matrix
$\mathbf{I}$, i.e., 
\begin{equation*}
I_{ij}(\mathbf{x}) = \begin{cases}
             1  & \text{if } \mathbf{x} \text{ is outlier with probability } \beta_{ij} \\
             0  & \text{Otherwise } 
       \end{cases}
\end{equation*} 
$\mathbf{I}$ is realized upon the immediate processing of
$\mathbf{x}$ and the available synopses and becomes 
the basis for the subsequent decision making.
Every $I_{ij}$ can be the outcome of any outliers detection 
technique (e.g., $\chi^{2}$, Grubb's test, clustering based) depending on the type of synopsis received by peers.
The simplest outliers detection technique is based 
on the identification of $\mathbf{x}$ as being `produced' by 
the distributions depicted by the statistics of 
every dataset $D_{i}$
\cite{kolomvatsosscalcom}.
Let this probability be $P(\mathbf{x}, D_{i})$.
In the group of $N$ EC nodes, we can consider  
$N \cdot M$ distributions, i.e., an individual distribution
for each dimension of $\mathbf{x}$.
Let $\Theta_{ij}, i=1, 2, \ldots, N \& j=1, 2, \ldots, M$ represent 
every distribution for the $j$th dimension in $D_{i}$ 
with a probability density function (pdf)
$f_{\Theta_{ij}}(x)$.
Hence, $P(\mathbf{x}, D_{i})$ can be defined 
as follows:
$P(\mathbf{x}, D_{i})= \prod_{\forall j} f_{\Theta_{ij}}(\mathbf{x})$.
Given the distributions $\Theta_{ij}$
and constant weights $w_{i} > 0$,
the pdf of the mixture 
is $f_{\Theta_{ij}}(\mathbf{x})=\sum_{\forall i} w_{i} \prod_{\forall j} f_{\Theta_{ij}}$. 

In general, $\mathbf{I}$ is a matrix hosting the outcomes of 
$V \cdot N$ Bernoulli trials with different success probabilities.
Our aim, before we decide that $\mathbf{x}$ is an outlier,
is to detect if multiple indicators agree upon this event
for multiple datasets.
Actually, we want to perform
a double majority voting,
i.e., the first per column (multiple indicators for the same dataset)
and the second per row (multiple aggregated indicators for the ecosystem of datasets).
For the envisioned double majority voting, we adopt 
the $\delta$-majority function upon 
$V$ binary variables \cite{crama}.
$\delta$ is the threshold over which we consider 
$\mathbf{x}$ as an outlier for a specific dataset.
The following equation holds true:
\begin{equation}
 B(I_{ij}, j) = 1 \iff \sum_{i=1}^{V} I_{ij} \geq \delta 
\end{equation}
where $B(\cdot)$ is the indicators aggregation function.
In the above equation, we can also incorporate the confidence that an outlier detectors has on the reported result. This way, we focus on a `fuzzy' outliers detection 
methodology that apart from the binary indication, every detector reports a real value which is the probability of
$\mathbf{x}$ to be an outlier or not.
In that case, the final outcome is delivered by the following equation
$B(I_{ij}, j) = 1 \iff \sum_{i=1}^{V} w_{i}I_{ij} \geq \delta$
where $w_{i}$ is the probability/confidence reported by the $i$th detector.
The second axis of aggregation is performed
upon the  $B(I_{ij}, j) = B_{j}$ realizations.
Again, a $\delta'$-majority function is adopted, i.e.,
\begin{equation}
 B'(B) = 1 \iff \sum_{j=1}^{N} B_{j} \geq \delta' 
\end{equation}
$B'$ represents the view of our model 
that $\mathbf{x}$ should be characterized as an outlier or not.

As $\mathbf{I}$ is the set of $V \cdot N $ Bernoulli trials with different success probabilities, 
we can easily deliver the final success probability
for any incoming data vector.
The sum of the outcomes of the aforementioned 
Bernoulli trials can be adopted to define the 
variable
$Z = \sum_{i=1}^{V} \sum_{j=1}^{N} I_{ij}$ which follows
a Poisson Binomial distribution with probability mass function
(pmf)
\begin{eqnarray}
	P(Z=z) = \sum_{A \in \mathcal{F}_{m}} \prod_{i \in A} \beta_{ij} \prod_{j\in A^{c}} (1-\beta_{ij})
\end{eqnarray}
with 
$\mathcal{A}$ being the set of all 
subsets of $m$ node indexes selected from 
$\left\lbrace 1, 2, \ldots, N \right\rbrace$ and 
$A^{c}$ is the complement of the set $A$. 
When all $\beta_{ij}$s are identical, $Z$ follows 
the binomial distribution. 
Now, we can easily define the success probability that 
$\mathbf{x}$ will be identified as outlier in the entire group of detectors and nodes.
Initially, we have:
The following equation holds true \cite{kolomvatsosscalcom}:
\begin{eqnarray}
	F(z) = \sum_{m=z}^{N} \left\lbrace \sum_{A \in \mathcal{F}_{m}} \prod_{i \in A} \beta_{ij} \prod_{j\in A^{c}} (1-\beta_{ij}) \right\rbrace
\end{eqnarray}
$F(z)$ indicates the probability 
of having at least $z$ outlier identification results out of $N$. 
In the above equation, the lowest value for $z$ is  
\begin{eqnarray}
	z = \left\{
\begin{array}{ll}
      \frac{N}{2} + 1 & \text{if $N$ is even} \\
     \frac{N+1}{2}  & \text{if $N$ is odd} \\
\end{array} 
\right.
\end{eqnarray}
When $N$ is high, it is not easy to 
calculate all the necessary subsets for 
$\mathcal{F}_{m}$, thus, we have to rely on a computationally efficient method, i.e.,
on one of the methodologies 
presented in \cite{hong}.
For instance, we could adopt an approximation model (e.g., Poisson or Normal)
to quickly deliver the final probability of having $\mathbf{x}$
as an outlier, thus, rejected from further local processing.
As far as the `individual' outlier identification process concerns, 
we can rely on widely adopted techniques,
i.e., statistical-based (parametric or non-parametric approaches), nearest neighbor-based, clustering-based, classification-based (Bayesian network-based and support vector machine-based approaches), and spectral decomposition-based approaches. 
The proposed model can incorporate any number of outliers detectors from the aforementioned categories.

\subsection{Statistical Management of Data Vectors}
\label{replication}
Assume that $\mathbf{x}$ arrives in the EC node $n_{i}$.
If $\mathbf{x}$ is not an outlier, $n_{i}$ should decide to replicate the vector
in a sub-set of nodes. 
Our target is to apply the replication process only for nodes
that are highly correlated with $\mathbf{x}$.
The replication of 
$\mathbf{x}$ in the entire ecosystem
will flood the network with messages in addition 
to the `disturbance' of the statistics of uncorrelated datasets
`invited' to host $\mathbf{x}$.
In our scheme, we select the top-$k$ similar datasets
where $\mathbf{x}$ will be replicated.
The parameter $k$ is lower than $N$ 
to reduce any negative effects in network communications.

Our replication process is realized upon 
the collected synopses $s_{i}, \forall i$ instead 
of relying on a voting process adopting two correlation metrics
like the model presented in \cite{kolomvatsosscalcom}.
In this paper, the novelty is that we detect the 
trend of the correlation between 
$\mathbf{x}$ and the available synopses $s_{i}, \forall i$
estimating the unknown distribution that depicts
their similarity. 
The latest $W$ synopses reports
$\left\lbrace s^{t}_{i} \right\rbrace_{t=1}^{W}$
become the basis for our mechanism.
We want to detect if $\mathbf{x}$
is similar to the latest 
$W$ `views' on each dataset.
We rely on the 
similarity between $\mathbf{x}$
and each $s_{i}$ depicted by the function
$g(\cdot)$.
Let $g(\cdot)$ be the distance function,
i.e., $g(\mathbf{x}, s_{i}) = \lVert \mathbf{x}, s_{i} \rVert$. 
For instance, if synopses are depicted by the mean of each 
dimension, $g(\cdot)$ can represent the like
the $L_{p}$ norm, $p = 1, 2, \ldots$ with the $\mathbf{x}$.
When $s_{i}$ is depicted by the centroids of a set of clusters, 
$g(\cdot)$ can be the distance with the closest centroid or the accumulated distance with the set of the reported centroids.
For simplicity in our notation, we consider $g_{i}$ as the 
distance calculated by $g(\mathbf{x}, s_{i})$.

Based on the collected synopses, we can have a time series of distances for $\mathbf{x}$, i.e.,
$g^{t}_{i}, t=1,2, \ldots, W$.
Upon these distances, we can expose their unknown pdf targeting 
to extract the probability of having the similarity 
between $\mathbf{x}$ and $s_{i}$ upon a pre-defined threshold.
We rely on the widely known Kernel density Estimation (KDE)
\cite{Baszczynska} to derive the pdf of $g_{i}$.
Based on $g^{t}_{i}$, we estimate $F_{G}(g_{i})$ via estimating the pdf $f_{G}(g_{i})$. 
Having $W$ recent samples for $g_{i}$,
$f_{G}(g_{i})$ is estimated as
$\hat{f}_{G}(g_{i}, W) = \frac{1}{W \cdot h} \sum_{t=1}^{W} K\left( \frac{|g_{i}-g^{W-t+1}_{i}|}{h} \right)$,
where $h > 0$ is the bandwidth of the symmetric kernel function $K\left(u\right)$ (integrating to unity). One of the most frequent adopted kernel function is the Gaussian, i.e.,
$K\left(u\right) = \frac{1}{\sqrt{2 \pi}} e^{-\frac{1}{2}u^{2}}$.
For saving time and alleviate the complexity of the proposed model, we rely on an incremental estimation of $\hat{f}_{G}$ n. 
The pdf $\hat{f}_{G}(g_{i}, t)$ for $t = 1, \ldots, W$ is incrementally 
estimated by its previous estimate $\hat{f}_{G}(g_{i}, t-1)$ and the current 
value $g^{t}_{i}$, that is, we recursively obtain for $t=1, \ldots, W$ that: 
\begin{equation} 
\hat{f}_{G}(g_{i}, t)  =\frac{t-1}{th}\hat{g}_{i}(g_{i}, t-1) + \frac{1}{th}K\left(\frac{|g_{i}-g^{W-t+1}_{i}|}{h}\right)
\label{eq:5}
\end{equation}
If we apply the Gaussian function on the KDE, we obtain an estimation of the cdf $\hat{F}_{G}(g_{i}, W) = \int_{g_{\min}}^{g_{i}}\hat{f}_{G}(u, W)du$ using the $W$ values $\{g_{i}^{W-t+1}\}_{t=1}^{W}$:
\begin{equation}
\label{eqx}
\hat{F}_{G}(g_{i},W) = \frac{1}{W} \sum_{t=1}^{W} \frac{1}{2} \left( 1 + \mbox{erf} \left(\frac{g_{i} - g_{i}^{W-t+1}}{\sqrt{2}} \right)\right) 
\end{equation}
where $\mbox{erf} \left( \cdot \right)$ is the error function. 
Hence, at time $t$, we obtain the estimation 
of $\gamma_{i}  = P(g_{i}^{t} > \epsilon) \approx 1 - \hat{F}_{G}(\epsilon, W)$.
In the first place of our future research agenda is the 
incorporation of an estimation model for multiple time steps forward than 
$W$ in the decision making mechanism.

After the calculation of $\gamma_{i}$, 
we rely on an ordered list 
of the available peer nodes. 
We rank peers upon their $\gamma_{i}$ result to conclude the 
optimal solution for the specific setup.
The \textit{Probability Ranking Principle} \cite{jones} dictates that if peers are 
ordered by decreasing $\gamma_{i}$ over the available datasets and 
$\mathbf{x}$, 
the effectiveness of the model is the best to be gotten for those 
instances.
From the ranked list, we select the top-$k$ outcomes and the 
corresponding nodes to host
$\mathbf{x}$.
Every selected $n_{j}$ should, then, 
update the corresponding statistics of their datasets.
We adopt a delay mechanism in the delivery 
of messages related to the new synopses taking
into consideration the difference between the new synopsis with the previous one and the remaining time till the expiration of the pre-defined interval when 
nodes should report their synopses. 
A time optimized process for delivering synopses while delaying the final 
decision is studied 
in \cite{kolomvatsostkde} and is not subject of the current 
effort.

\section{Experimental Evaluation}
\label{setup}

\subsection{Performance Metrics \& Setup}
The evaluation of the proposed model involves 
a set of experimental scenarios upon a real trace.
The aim is to investigate the performance
of our mechanism concerning the 
number of the detected outliers as well as
the number of the stored data vectors that deviate 
from the statistics of every dataset.
We also focus on the evaluation of the proposed model concerning its 
ability of replicating 
the incoming data 
to the appropriate datasets keeping their
solidity at high levels.
Our simulations involve a high number of data vectors generated 
in various nodes into the network.
When a data vector is produced, 
we adopt different distributions
for producing the corresponding values for each dimension.
Our dataset is retrieved by \cite{devito}.
It contains 9358 instances of hourly averaged responses from 
an array of five (5) metal oxide chemical sensors embedded in an Air Quality 
Chemical Multisensor Device. 
This device was placed in a highly polluted area in an Italian city. 
All values in the dataset are normalized in the unity interval.

The performance of the proposed mechanism is evaluated by a set of metrics as follows:
\textbf{(i)} the percentage of the detected outliers $\omega$.
As the adopted dataset does not contain outlier vectors, we
randomly `produce' fake outliers by changing the values 
of the dataset. The aim is to identify if the proposed ensemble outliers detection 
scheme is capable of rejecting vectors that may jeopardize the solidity 
of datasets. We have to notice that 
$\omega$ refers in the detection of vectors that are considered 
as outliers for the entire ecosystem. For this,
we incorporate into our ensemble outliers detection model with 
following individual detection methods: (i)
the probabilistic approach discussed in Section \ref{mechanism}, i.e.,
we consider the probability of the incoming vector to be 
produced by the distribution characterizing every dimension of the available datasets; (ii)
a statistical approach, i.e., we detect if the incoming 
vector significantly deviates from the mean of each dimension; and (iii)
the $\chi^{2}$ metric.
In our experimental evaluation, we produce a number of outliers equal to 
1\% of the retrieved values;
\textbf{(ii)} the percentage of data vectors that deviate from the statistics of each
dataset $\tau$. 
$\tau$ depicts the average number of `local outliers'
stored in every dataset after the application of our processing model compared to the 
total number of vectors. 
As a `local outlier', we define 
the data vector deviating three times the standard deviation from the mean (under the assumption that data follow a Gaussian distribution).
We try to detect if the
proposed approach is capable of storing similar data vectors to the same datasets. 
When $\tau$ is high means that multiple vectors do not match with the remaining vectors in the dataset. 
When $\tau \to 0$, our datasets do not contain any `outlier' data;
\textbf{(ii)} the solidity of the formulated
datasets as depicted by the mean $\mu$ and
standard deviation $\sigma$. 
We target to a low $\sigma$, i.e., to deliver solid datasets.
When $\sigma$ is low means that the datasets are concentrated around the mean, thus, we have a clear view on their dispersion. Such a result, as mentioned above, can assist in the efficient assignments of queries into the appropriate datasets. 
We perform a set of experiments for different $N$, $k$, $M$ and $W$ taking their values
as follows: $N \in \left\lbrace 10, 50, 100 \right\rbrace$, $k \in \left\lbrace 2, 5 \right\rbrace$,
$M \in \left\lbrace 2, 10 \right\rbrace$ and
$W \in \left\lbrace 10, 50 \right\rbrace$.
In total, we consider that 1,000 data vectors are received by the group of nodes
and report our results for the aforementioned metrics.

\subsection{Performance Assessment}
\label{assessment}
In Tables \ref{tableres1} ($W=10$) \& \ref{tableres2} ($W=50$), we present our results for $\omega$ and $\tau$ metrics.
We observe that the proposed model is capable of detecting the generated outliers ($\omega \in [0.70, 1.0]$) while keeping similar data to the same datasets especially when 
$N$ is low. 
$\omega$ is generally retrieved to be equal to 
unity except when the number of EC nodes is low and the number of dimensions is high.  
Additionally, an increased $k$ leads to an increased 
$\tau$ due the fact that we replicate the incoming 
data vectors to multiple EC nodes. 
For instance, if $W=10$, we observe that 
the number of `local outliers' are 0.5\% when 
$N=10$ and reach 30\% \& 40\% when $N =100$.
This means that in the scenario where we consider a high number of nodes, the dispersion of datasets increases.
This situation is also affected by the increased number of nodes selected to replicate every accepted data vector ($k=5$).
In case where $W=50$, we get similar results that make us understanding that the size of the adopted window for performing the KDE processing does not affect the performance of the proposed model. 

\begin{table}
\centering
\caption{Performance outcomes for various experimental scenarios and $W=10$}
\label{tableres1}
\begin{tabular}{c|cc|cc|cc|cc}
\hline\hline
\multicolumn{1}{c|}{} & \multicolumn{4}{c|}{$k=2$}                                                 & \multicolumn{4}{c}{$k=5$}                                                                          \\
                     & \multicolumn{2}{c|}{$M=2$}                         & \multicolumn{2}{c|}{$M=10$} & \multicolumn{2}{c|}{$M=2$} & \multicolumn{2}{c}{$M=10$}  \\
$N$   & $\omega$ & $\tau$ & $\omega$ & $\tau$ & $\omega$ & $\tau$ & $\omega$ & $\tau$  \\
\hline\hline
10   & 1.00  & 0.005  & 0.90  & 0.10  & 1.00   & 0.005 & 0.85  & 0.11  \\
50   & 1.00  & 0.08   & 1.00  & 0.28  & 1.00   & 0.10  & 1.00  & 0.30   \\
100  & 1.00  & 0.10   & 1.00  & 0.30  & 1.00   & 0.20  & 1.00  & 0.40     \\
\hline\hline               
\end{tabular}
\end{table}

\begin{table}
\centering
\caption{Performance outcomes for various experimental scenarios and $W=50$}
\label{tableres2}
\begin{tabular}{c|cc|cc|cc|cc}
\hline\hline
\multicolumn{1}{c|}{} & \multicolumn{4}{c|}{$k=2$}                                                 & \multicolumn{4}{c}{$k=5$}                                                                          \\
  & \multicolumn{2}{c|}{$M=2$}  & \multicolumn{2}{c|}{$M=10$} & \multicolumn{2}{c|}{$M=2$} & \multicolumn{2}{c}{$M=10$}  \\
$N$   & $\omega$ & $\tau$ & $\omega$ & $\tau$ & $\omega$ & $\tau$ & $\omega$ & $\tau$  \\
\hline\hline
10   & 0.87  & 0.002  & 0.70  & 0.004  & 0.95   & 0.05 & 0.70  & 0.02  \\
50   & 1.00  & 0.10   & 1.00  & 0.10  & 1.00   & 0.06  & 1.00  & 0.12   \\
100  & 1.00  & 0.20   & 1.00  & 0.30  & 1.00   & 0.18  & 1.00  & 0.33     \\
\hline\hline               
\end{tabular}
\end{table}

\begin{figure}[H]
       \centerline{\includegraphics[width=10cm,height=7cm]{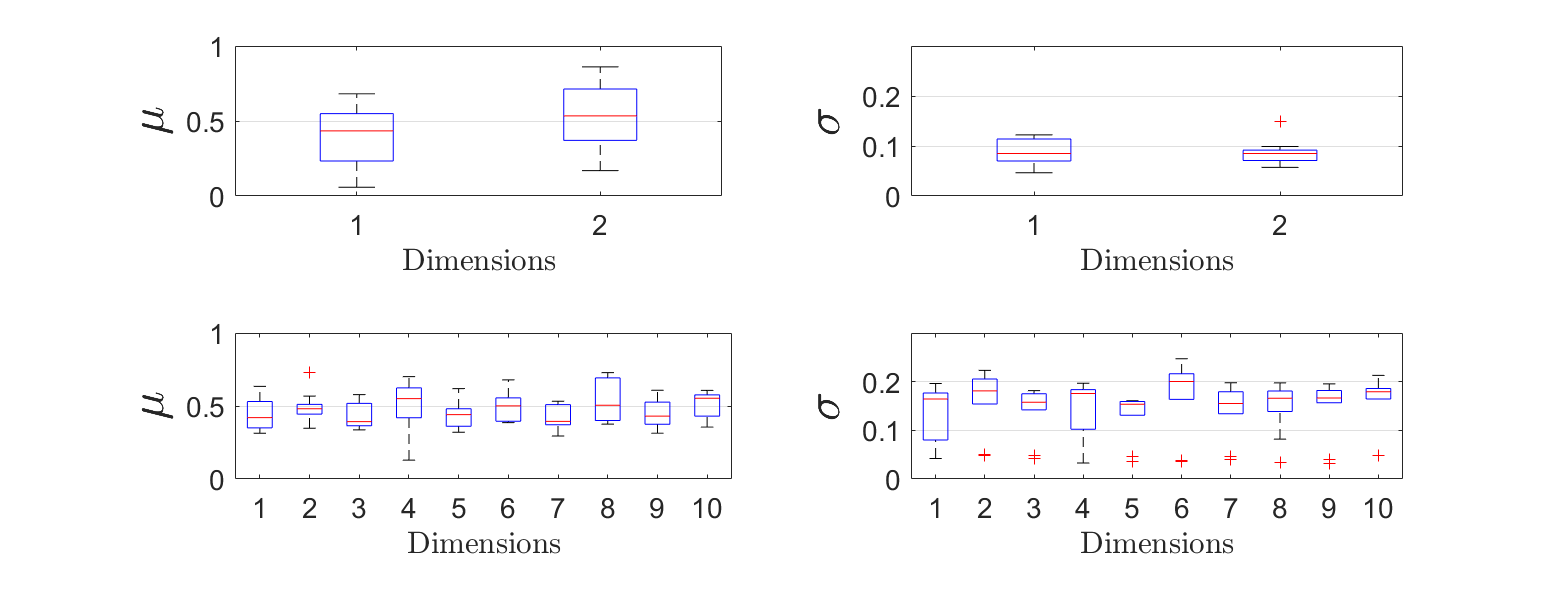}}
	\caption{Performance outcomes related to the solidity of the retrieved datasets for $N=10$ and $k=2$ (up: $M=2$, down: $M=10$)}
	\label{fig:figMean1}
\end{figure}

In Figure \ref{fig:figMean1}, we provide experimental outcomes for $N=10$ taking into consideration the mean and the standard deviation of the retrieved datasets.
We conform our findings as discussed above, i.e., 
the number of dimensions affect the dispersion of data. We observe that $\sigma$ is low when $M$ is low as well.
We also observe the local outliers in case where 
multiple dimensions are adopted 
by our model. 
We conclude that the aggregation of the difference 
between the available synopses and data vectors when 
$M$ is high may jeopardise the solidity of datasets.
However, the majority of $\sigma$ realizations is retrieved below the median for a high number of datasets. 

\begin{figure}[H]
       \centerline{\includegraphics[width=10cm,height=7cm]{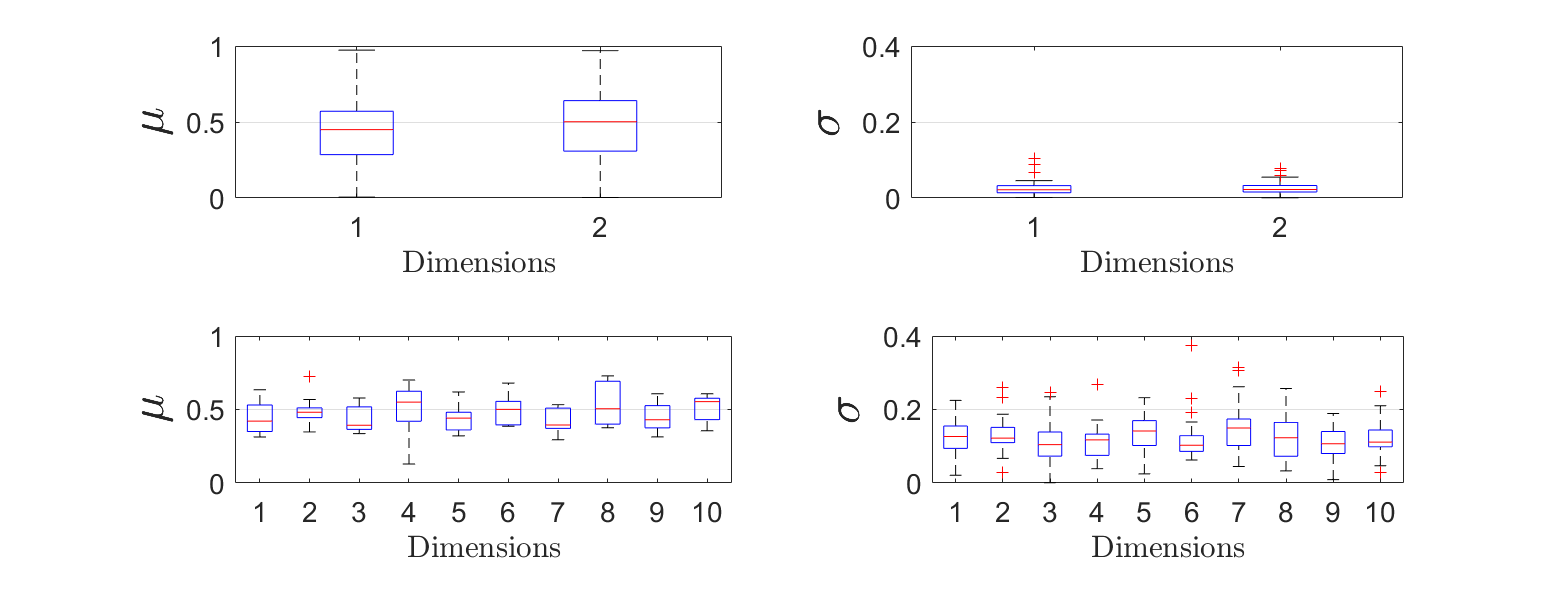}}
	\caption{Performance outcomes related to the solidity of the retrieved datasets for $N=100$ and $k=2$ (up: $M=2$, down: $M=10$)}
	\label{fig:figMean2}
\end{figure}

\begin{figure}[H]
       \centerline{\includegraphics[width=10cm,height=7cm]{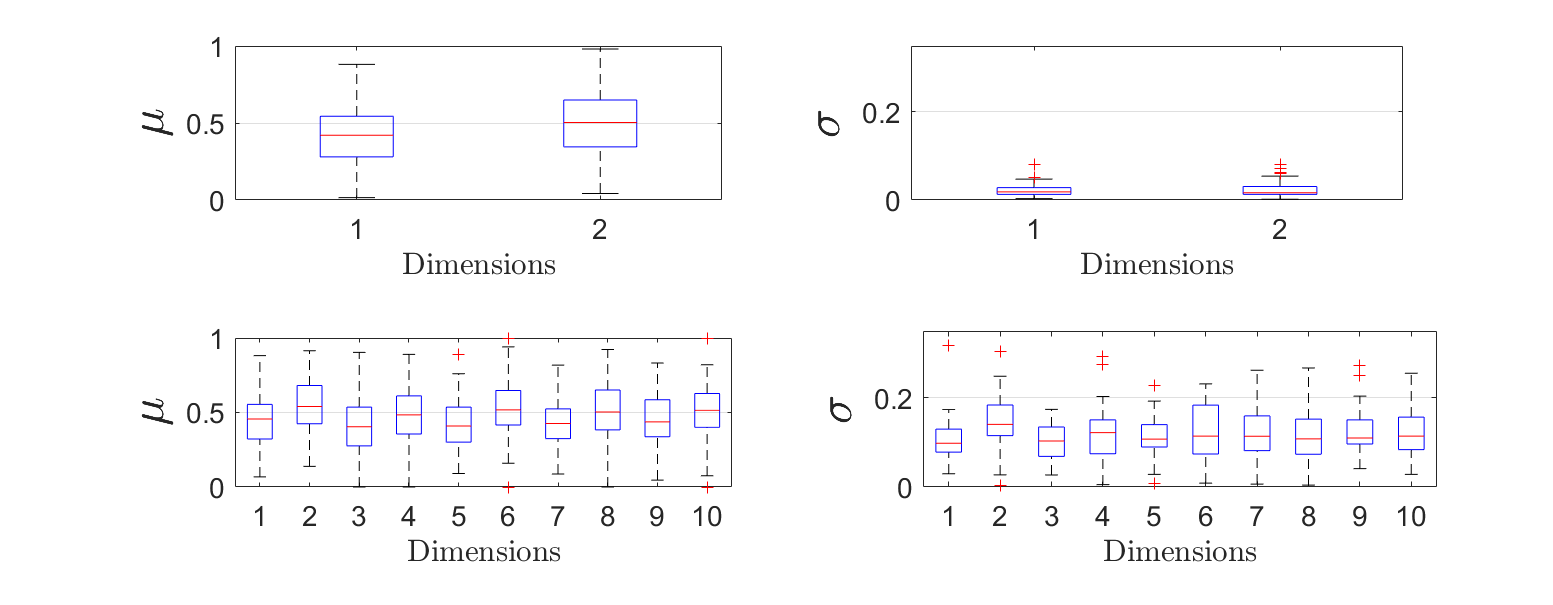}}
	\caption{Performance outcomes related to the solidity of the retrieved datasets for $N=100$ and $k=5$ (up: $M=2$, down: $M=10$)}
	\label{fig:figMean3}
\end{figure}

In Figure \ref{fig:figMean2}, we present our performance outcomes for $N=100$.
A low $M$ leads to a very low $\sigma$. Again, an increased $M$ may lead to an increased $\sigma$ as well.
Now, $\sigma$ is lower than in the previous experimental scenario when $M=10$, however, we can observe some outlier values over the maximum 
$\sigma$ realization.
Similar outcomes are retrieved 
in the scenario where
$k=5$ (see Figure \ref{fig:figMean3}).

The number of messages sent to the network 
depends on $k$. The lower the $k$ is the lower the number of the required messages.
In our experiments, the total number of messages are, in average, 1980 ($k=2$) or 
4940 ($k=5$). If we adopt the baseline model for replication purposes, i.e., we replicate 
every accepted data vector to the entire network, we require 9900 messages when 
$N=10$ to 99000 when $n=100$ (in average).
Finally, the time required for delivering the final result (in seconds)
is in the interval [0.004, 0.0006] when $N \in \left\lbrace 10, 50, 100 \right\rbrace$.
The lowest value in the aforementioned interval is retrieved 
when all the adopted parameters get their lowest realizations and the 
maximum value of the interval corresponds to the maximum value 
for all parameters. 
The retrieved average time requirements refer in every incoming data vector and
depicts the ability of the proposed mechanism to deliver the final outcome in real time.
In average, our model can serve, approximately, 160 to 250 data vectors per second.

\section{Conclusion}
\label{conclusions}
We discus a simple, however, efficient
probabilistic model for allocating data 
collected by IoT nodes to a set of EC nodes.
We propose the use of an ensemble model for outliers detection
in the pre-processing phase and the probabilistic allocation
upon data synopses reported by EC nodes.
The aim is to identify the datasets that match to the 
incoming vectors in order to support 
an efficient replication process.
The proposed decision making is applied upon historical synopses 
creating a time series of similarity/difference realizations with the 
incoming data vectors. 
This way, we are able to build a fault tolerant EC infrastructure
where nodes could perform the processing
of numerous tasks/queries.
The proposed mechanism manages to conclude the allocation for each incoming 
vector in real time.
This is a strategic decision as we adopt techniques that 
demand limited time to provide the final results.
Our performance outcomes upon a real trace indicate 
the advantages of the proposed scheme. 
The outliers detection rate is optimal for the vast majority of the experimental scenarios while the solidity of the formulated datasets is kept at high levels.
In the first places of our future research agenda is to incorporate a communication model between EC nodes and include the aspects of that model into the decision making process.
This will increase the applicability of the proposed mechanism being aligned with real needs. 


\end{document}